\documentclass{elsart}

\input{psfig.sty}

\journal{Physics Letters A}

\begin{document}

\begin{frontmatter}

\title{Experimental multi-photon-resolving detector
using a single avalanche photodiode}

\author[adr1,adr2]{Ond\v{r}ej Haderka}
\ead{haderka@sloup.upol.cz}
\author[adr2]{Martin Hamar}
\author[adr1,adr2]{Jan Pe\v{r}ina Jr.}

\address[adr1]{Joint Laboratory of
Optics, Palack\'{y} University and Institute of Physics of
Academy of Sciences of the Czech Republic, 17. listopadu 50,
772 07 Olomouc, Czech Republic}
\address[adr2]{Department of Optics,
Palack\'{y} University, 17. listopadu 50, 772 07 Olomouc,
Czech Republic}

\begin{abstract}
A~multichannel detector has been constructed using a single
avalanche photodiode and a fiber-loop delay line. Detection
probabilities of the channels can be set using a variable-ratio
coupler. The performance of the detector is demonstrated on
its capability to distinguish multi-photon states
(containing two or more photons) from the one-photon state and
the vacuum state.
\end{abstract}


\end{frontmatter}

\section{Introduction}

There is an increasing need for identifying the number of photons
that is contained within a weak light pulse or time interval of a
weak cw light field. A~device capable of photon-number resolution
would contribute both to fundamental research in the area of
quantum optics and to more-or-less practical quantum communication
systems, such as quantum key distribution schemes \cite{gisin}. In
the former case it would significantly help in the analysis and
preparation of quantum states with prescribed statistics
\cite{kok}, in the latter it would prolong communication distances
and increase the rate of secure communication
\cite{brassard,my,walton}.
A detector capable to resolve the vacuum state, one-photon Fock
state and states containing more than one photon
was recently constructed \cite{kim,brattke}, but it requires
operation under
extreme conditions at present and it hardly becomes a common
laboratory tool at least in near future.
A similar problem occurs with a photon-number resolving detector
\cite{miller}
based on superconducting transition-edge sensor microcalorimeter
technology that needs mK temperatures for its operation.

Common detectors of weak light fields (avalanche photodiodes and
photomultipliers) cannot provide photon-number resolution but they
have high quantum efficiencies in the visible \cite{cova} and near
infrared \cite{ribordy} regions. The use of a cascade of such
detectors behind a $1\times N$ multiport \cite{jmo,kok} then
provides a device capable to resolve photon-numbers to some
extent. Assuming a lossless device, ideal detectors, and provided
that the mean number of photons in the signal $\mu \ll N$, a part
of any multi-photon signal containing $k$ photons gets split with
a high probability to the arms of the multiport in such a way that
$k$ detections at different detectors occur. (We note that in the
limit $ N \rightarrow \infty $ this probability is one.) However,
it was shown in \cite{kok} that the performance of such a device
is severely inflicted by losses in the device and imperfections of
the detectors. To achieve a reasonable performance, large array of
detectors with high quantum efficiency and low noise is required
\cite{kok}, so that such photon-number resolving device becomes
unacceptably complex. The use of this photon-number resolving
device in a source of one-photon Fock states based on entangled
photon pairs and postselection was analyzed in \cite{my}. It was
shown that photon fields with the Fano factor around 0.7 can be
generated by this source under real conditions. A quantum-key
distribution system using this source can have a secure
communication distance up to 120~km and can provide higher values
of gain \cite{my}.
These results motivate the endeavour to simplify the above
discussed photon-number resolving device.
We note that also single-photon sources utilizing NV centers
\cite{kurtsiefer} and quantum dots \cite{santori,pelton,stevenson}
are perspective for quantum-key distribution systems.
However, NV centers have low efficiencies of single-photon
generation ($ \approx 10^{-3} $) and quantum-dot sources
have to be cooled to 5~K at present.

In this letter we propose and test a variant of such
cascading device in which we replace the $1\times N$ multiport and $N$
detectors with a fiber-loop delay line and a single
avalanche-photodiode detector. This decreases the complexity as
well as the cost of the
device to a reasonable measure. A~similar device has been recently
suggested also by other authors \cite{banaszek}. The number of
detectors $N$ is given by the number of time windows (channels) we
detect using the time-of-flight spectrometer. We use
a variable-ratio coupler at the entrance to the fiber-loop delay
line so that we can control the distribution of probabilities of
detection in the individual channels.
In order to show the performance of this fiber-loop detection device,
we theoretically determine and
experimentally test the best setting of the
device with respect to distinguishing multi-photon states
(containing two or more photons) from the one-photon state and
the vacuum state.

\section{Description of the device}

\begin{figure}
 \centerline{\psfig{file=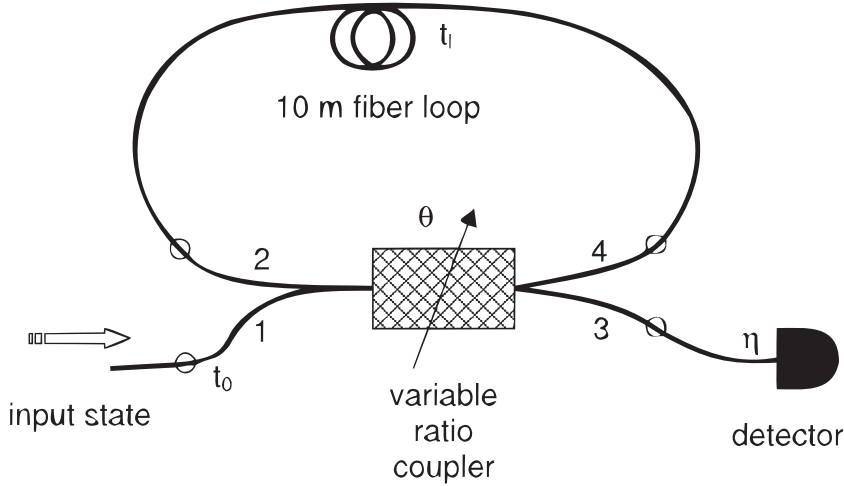,width=0.8\textwidth}}

  \caption{Scheme of the device. Incoming quantum state is fed
  to input 1 of the variable
  ratio coupler (SVR). Detector is connected to pigtail 3. Pigtails 2
  and 4 are connected with a 10 m long fiber patchcord that serves
  as a delay line. All pigtails of the SVR are 1 m long. Circles denote
  FC connectors.}
  \label{fig1}
\end{figure}

The scheme of our device is plotted in Fig.~\ref{fig1}. Quantum
state to be analyzed is injected to port 1 of the single-mode
variable ratio coupler (SIFAM SVR-82, referred as SVR hereafter).
Port 3 outputs to a detector module based on a silicon avalanche
photodiode with active quenching (Perkin-Elmer SPCM-AQ-141-FC).
This detector yields TTL output upon incident photon with quantum
efficiency $\eta$ of 60\% (at 830~nm, including coupling-optics
losses) and is not capable of
resolving the number of incident photons. The dark count rate is
about 40 counts per second. Ports 2 and 4 are interconnected with
10~m long single-mode fiber thus forming a delay line longer than
the dead time of the detector which is 50~ns. The outputs of the
detector are registered as stop pulses in the time-of-flight
spectrometer (Fast ComTec TOF7885) that is connected to a PC
directly or through a multichannel buffer. Start pulses for the
time-of-flight spectrometer are generated by trigger pulses of the
source of quantum states to be examined.

Denoting the intensity transmission coefficients from port $i$ to
port $j$ of the SVR by $t_{ij}$, the device input coupling
transmission by $t_0$ (due to loss at the input connector), the
SVR transmission by $\theta$ (due to SVR excess loss), the total
transmission of the fiber loop by $t_l$ and the transmission from
port 3 of the SVR to the detector including detector quantum
efficiency by $\eta$, we arrive at the transmission
coefficients $ h_1, h_2, \ldots $ for the detection channels:
\begin{eqnarray}
  h_1 &=& t_0 \theta t_{13} \eta, \label{probgena}\\
  h_k &=& t_0 t_{14} \theta^k t_l^{k-1} t_{23} t_{24}^{k-2} \eta,
  k\geq2. \label{probgenb}
\end{eqnarray}
If a single photon enters the device, transmission coefficients give
also the probabilities of photon detection.

A~question arises, what is the optimum setting of the SVR for the
detection of multi-photon states. The highest probability to
distinguish multi-photon states from the one-photon state and the
vacuum state is achieved provided that the probabilities $ h_1,
\ldots, h_N  $ have the same value. The reason is that the energy
in a detected photon field is equally distributed over all
detectors and then the intensities of the photon fields in all
detectors reach the lowest possible value. This case can be also
distinguished by maximizing Shannon entropy. We cannot reach
this best case with our fiber-loop detection device. Best
conditions from the point of view of photon-number resolution
obtainable by our fiber-loop detection device can be found only
numerically in principle. However, numerical calculations as well
as experimental results indicated that the principle of
maximization of Shannon entropy is valid for this task and
can provide these conditions (see Fig.~4).

We first simplify the
real SVR to an idealized unitary device setting $\theta=t_0=1$,
$t_{13}+t_{14}=1$, $t_{23}+t_{24}=1$, and $t_{13}+t_{23}=1$. This
idealized coupler can then be represented with a single
variable-division-ratio parameter $r$. Upon replacing
$t_{13}=t_{24}\rightarrow r$ and $t_{14}=t_{23}\rightarrow (1-r)$
we get:
\begin{eqnarray}
  h_1 &=& \eta r, \label{probideala}\\
  h_k &=& \eta (1-r)^2 t_l^{k-1} r^{k-2}, k\geq2. \label{probidealb}
\end{eqnarray}
Assuming further $t_l=\eta=1$ in the ideal case, we can evaluate
Shannon entropy
\begin{equation}\label{entropy}
  E=-\sum_i h_i \ln(h_i)
\end{equation}
of the ideal channel detector as follows:
\begin{equation}\label{entropyideal}
  E=-2r \ln(r)-2(1-r) \ln(1-r).
\end{equation}
This entropy is maximized for $r=1/2$, i.e., a balanced SVR is
optimal. In the more general case $0<t_l,\eta, \theta<1$ the
condition for maximum entropy needs to be evaluated numerically
and it can be found that for realistic values of $t_0, t_l, \eta,
$ and $ \theta$ the maximum value of entropy is decreased and
the maximum occurs
for $r<1/2$. Thus, in a real device we can expect best performance
of the detector when SVR is unbalanced in favor of port 4. For the
parameters of our device we get the maximum value of entropy at $r=0.446$.

\section{Experimental results}

For an experimental test of the detector we have fed it with a
source of faint laser pulses with Poissonian statistics and
variable mean photon number. They have been obtained from a laser
diode (SHARP LT015) yielding 4 ns pulses that have been
subsequently attenuated by a digital variable attenuator (OZ
Optics DA-100) to a single-photon level.

\begin{figure}
 \centerline{\psfig{file=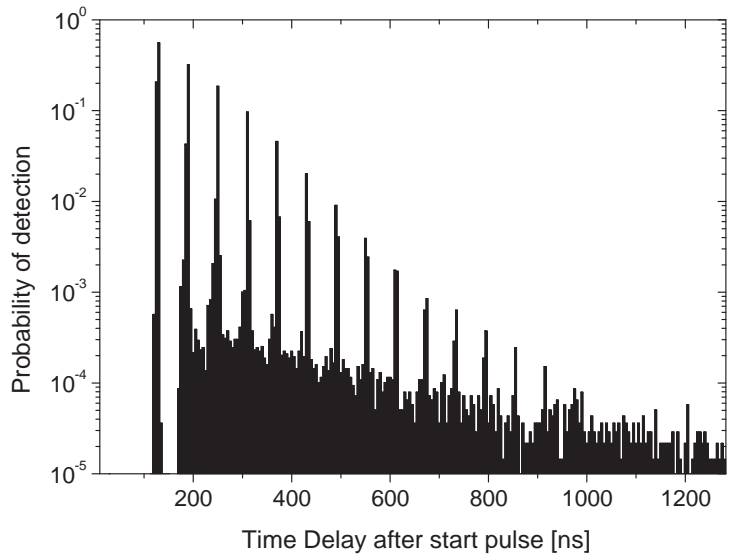,width=0.8\textwidth}}
  \caption{\label{fig2}A typical time spectrum obtained using the
  time-of-flight spectrometer at the mean photon number
  $\mu \doteq 2.13$.
  The spectrum is shown as a histogram of probabilities of detection
  per time bin; 1024 bins by 5 ns have been registered. Please note
  the logarithmic scale.}
\end{figure}

A~typical time spectrum measured by the time-of-flight
spectrometer at the mean photon number $\mu \doteq 2.13$ (see
Eq.~(\ref{definition}) below for the definition of $ \mu $) is
shown in Fig.~\ref{fig2}. Multiple peaks can be resolved at time
distances of about 60 ns. The background between the peaks is
caused by afterpulses, i.e. false detections that occur after dead
time of the detector. This is why they are not present between the
first two peaks. In our detector, the total afterpulse probability
is of the order of $p_{ap}^{peak}\approx 8 \times 10^{-3}$.
The envelope of the afterpulse
probabilities coming from individual peaks then forms the
background pattern visible in Fig.~\ref{fig2}. At low intensities
the probability of detection due to an afterpulse is considerably
lower so that the main source of noise are the dark counts of the
detector whose probability is $2\times10^{-7}$ per 5 ns bin.

It is important to know the losses in the device. The total
transmission of the device may be evaluated using
Eqs.~(\ref{probgena}) and (\ref{probgenb}) as
\begin{equation}\label{total}
  T = \sum_{k=1}^\infty h_k =
  \eta t_0 \left[ \frac{\theta (t_{13} t_{24}-t_{14} t_{23} )}{t_{24}} -
  \frac{t_{14} t_{23} \theta}{t_{24} (t_{l} t_{24} \theta - 1)}
  \right].
\end{equation}
The quantity $T/\eta$ can be measured directly  and we have found
$T/\eta=0.78\pm 0.01$. Since $\eta$ is specified by the
manufacturer of the detector, the total transmission $T$ of the
device is known. However, upon inspecting Eq.~(\ref{total}) we can get
more insight into the structure of the losses.

We have observed experimentally that the total transmission $T$
depends only weakly on the SVR setting. We therefore again replace
the four $t_{ij}$ coefficients by a single parameter $r$ and
arrive at a simplified expression
\begin{equation}\label{totalsimp}
  T \approx \frac{\eta t_0 (2 t_l \theta -1)}{t_l} -
  \frac{\eta t_0 (t_l \theta -1)^2}{t_l (r t_l \theta -1)}.
\end{equation}
Since the dependence of $T$ on $r$ is weak, we
may deduct that $T$ is given dominantly by the first term in
Eq.~(\ref{totalsimp}). The value of $\theta$ was found
in an independent measurement to be $\theta=0.955$.
The transmission of the fiber loop $t_l$ can be
found from the measured values of the normalized channel
probabilities $H_k=h_k/T $ ($ \sum_{k=1}^{\infty} H_k=1$). In
particular, the ratio $H_{k+1}/(H_k H_1)$ evaluates to
\begin{equation}\label{ratioh}
\frac{H_{k+1}}{H_k H_1} \approx 2 \theta t_l -1,
\end{equation}
where only the first term of Eq.~(\ref{totalsimp}) has been used. From
experimental results we find the value $H_{k+1}/(H_k H_1)\approx
0.80$ to be almost independent of $r$ and $k$ (thus justifying the
approximations used). From Eq.~(\ref{ratioh}) we then get $t_l \approx
0.94$ (0.27~dB). Finally, the value of input coupling
transmission $t_0$ is evaluated from Eq.~(\ref{totalsimp}) as $t_0
\approx 0.92$ (0.36 dB). These values are rather low and are
caused mainly by the wear-out of the connectors used in our
laboratory setup. The application of new connectors or fused
fibers would further improve the performance of the device.

\begin{figure}
 \centerline{\psfig{file=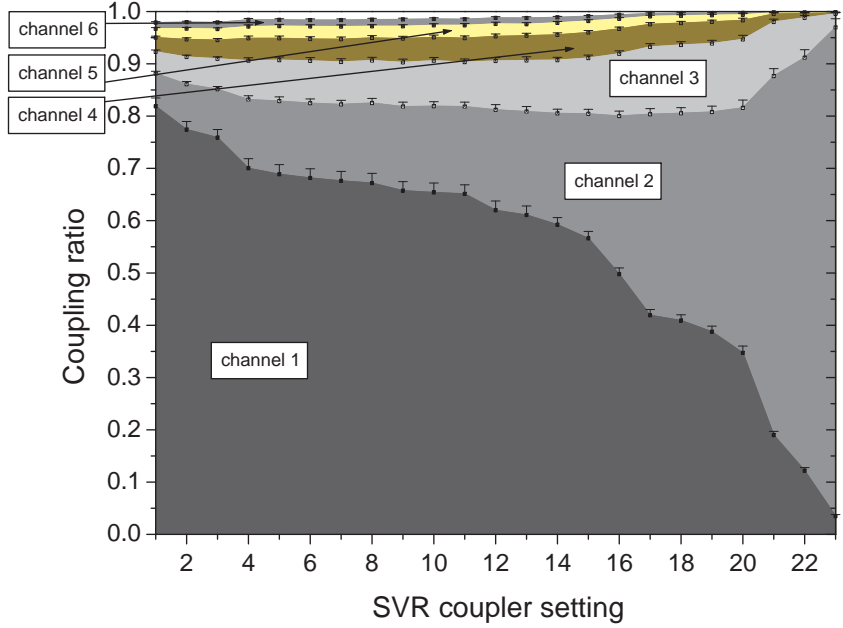,width=0.8\textwidth}}
  \caption{\label{fig3}Division ratio of the first six channels
  of the multichannel detector based on the SVR setting.
  The tick labels at x-axis roughly correspond to the scale of the
  micrometer screw.}
\end{figure}

The division ratio of the SVR is set by a micrometer screw. The
distribution of detection probabilities to the first six channels
based on the SVR position is shown in Fig.~\ref{fig3}. The small
white area in the top part of the plot corresponds to the sum of
the higher channels ($k>6$).
We can see that various settings are
possible. We can, e.g., set the device to the regime of one
dominant channel with other channels weak but a moderate number of
them (5 or 6) nonnegligible (their weight is above 1~\%; see positions
5-10 in Fig.~\ref{fig3}), or to a different regime of fewer but
relatively balanced channels (see, e.g., position 19 in
Fig.~\ref{fig3} where the first three channels are in the ratio
39\%:42\%:13\%).

\begin{figure}
 \centerline{\psfig{file=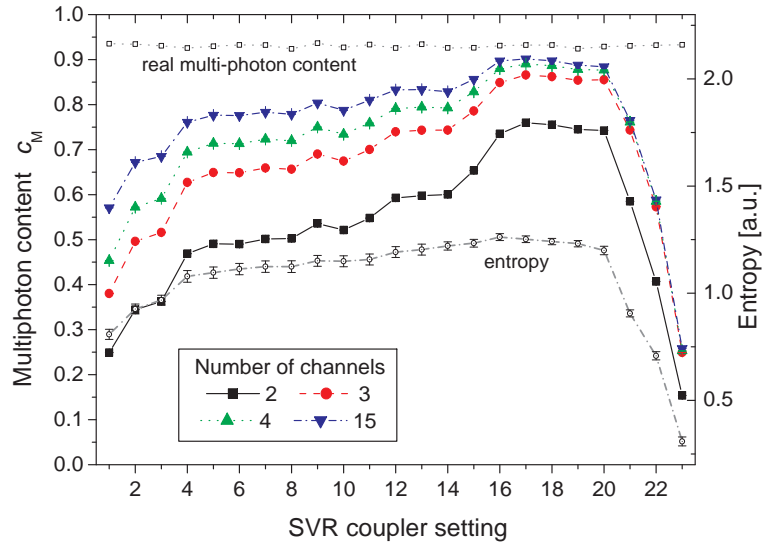,width=0.8\textwidth}}
  \caption{\label{fig4}Dependence of the multi-photon content $c_M$
  on the SVR setting when detected 2, 3, 4 or 15 channels of the device
  (see inset legend). A~real multi-photon content in the coherent state
  in front of the device (dotted line with open rectangles) is shown
  for comparison. The mean pulse energy $ \mu $ of the input state was 4.26
  photons per pulse. The dash-dot line with open circles is
  Shannon entropy
  computed from measured division ratios of the first 15 channels.}
\end{figure}

Let us now have a look at the capability to resolve multi-photon
states using our device. We characterize the measured Poissonian
signal by the mean photon number $ \mu $. The mean photon number $
\mu $ is obtained from the measured probability of detection $ p $
assuming detection of the whole Poissonian signal by the detector
with efficiency $T$, for which $ p = \sum_{i=1}^{\infty} p_n
= 1-p_0 = 1 - \exp(- T \mu ) $, i.e.
\begin{equation}
 \mu = - \ln (1-p)/ T .
 \label{definition}
\end{equation}
Our measured signal had $ \mu $  equal to 4.26 photons per pulse.
The probability of vacuum state detection $p_0$, single-channel
detection $p_1$, and multichannel detection $p_M=\sum_{k=2}^\infty
p_k$ were extracted from the measured data. At each setting, the
probabilities were obtained by detecting large number of laser
pulses ($\sim 10^5$). We characterize the fraction of multi-photon
states detected by our detector with another quantity, namely the
multi-photon content $ c_M $, $c_M=p_M/(p_1+p_M)$ ($ c_M $ gives
the probability that non-vacuum pulses contain more than 1 photon)
\cite{my}. Figure~\ref{fig4} shows the detected multi-photon
content $c_M$ at the mean photon number $\mu=4.26$ photons per
pulse. For comparison, a real (theoretical) multi-photon content
$c_M$ in a Poissonian laser pulse in front of the device is shown
as well ($c_M$ is not constant for different SVR settings because
the laser intensity slightly changed from one setting to another).
The highest values of $c_M$ have been observed for SVR positions
16-20 with maximum at position 17. The value of $c_M$ grows with
the number of channels considered. Due to losses in the device and
limited quantum efficiency of the detector, the measured
multi-photon content is lower than the real (theoretical)
multi-photon content. Nevertheless, the value of the measured
multi-photon content was lower by less than 4~\% in comparison
with the value of the real (theoretical) multi-photon content at
best performance of the device (SVR position 17, 15 channels) for
this pulse-energy level. It is also worth noting that the use of a
large number of channels is useful mainly in the regime with the
first dominant channel while it brings only a little advantage in
case of fewer but relatively balanced channels. Values of entropy
for each setting of the SVR are also plotted in Fig.~\ref{fig4}
and they clearly show that entropy of the multichannel detector is
the right indication of the multi-photon-resolution capability.
The optimum performance of the device is achieved close to the
value $r=0.453$ as estimated theoretically in the preceding
section.

\begin{figure}
 \centerline{\psfig{file=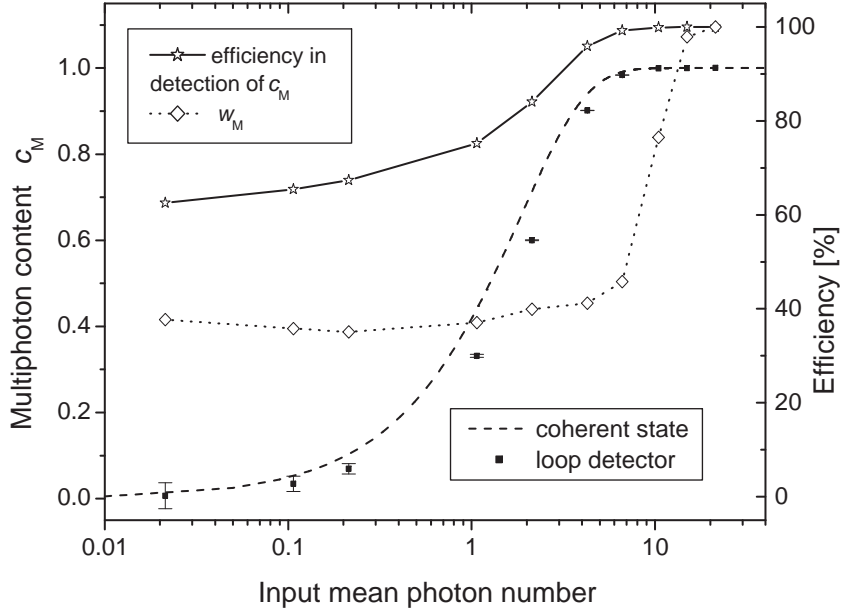,width=0.8\textwidth}}
  \caption{\label{fig5} Dependence of the measured multi-photon content
  $c_M$ (black rectangles) on the mean photon number $ \mu $ of the input
  coherent state when detected with 15 channels of the device. The
  real (theoretical) multi-photon
  content $ c_M $ of the input state is shown by the dashed curve.
  Solid curve with stars (right axis) gives the ratio of the measured
  value of $ c_M $ to the real (theoretical) one.
  The dotted curve with hollow diamonds (right axis) gives the ratio
  $ w_M $ (for the definition, see text below).}
\end{figure}

The performance of the
detector in identification of the multi-photon content over a wide
range of the mean photon number $ \mu $ of the input Poissonian state
is shown in Fig.~\ref{fig5}. The measured
values of the multi-photon content $c_M$ (black rectangles) follow
closely the (dashed) curve of the real (theoretical)
multi-photon content of the input
state. The ratio of the measured values of the multi-photon
content $c_M$ to the real (theoretical) ones shown in
Fig.~\ref{fig5} determines the efficiency of identification of
multi-photon states. It can be seen in Fig.~\ref{fig5} that
the ratio exceeds 60\% for weak coherent states and reaches almost
unity for strong coherent states.

Reduction of the multi-photon content $ c_{M,\rm in} $ of
a beam can be reached in the
following scheme. A source provides perfectly correlated
photon pairs (e.g., such pairs are generated by spontaneous
parametric downconversion). One of the correlated beams is postselected
by the measurement on the other beam using our fiber-loop detection
device. Postselection process provides a beam with lower values of
the multi-photon content $ c_{M,\rm out} $. The ratio $ w_M $
($ w_M = c_{M,\rm out} / c_{M,\rm in} $) then determines efficiency
of this reduction.
The dependence of the ratio $ w_M $ on the mean photon number $ \mu $
in Fig.~\ref{fig5} shows that the value of the ratio $ w_M $
can be reduced down to 40~\% for input Poissonian states with
$ \mu \le 5 $.
The most interesting input states for applications utilizing
multi-photon reduction are those
containing a more-or-less balanced mixture of vacuum,
single-photon and multi-photon contributions.
The reason is that weak
Poissonian states contain only a small fraction of multi-photon states
(e.g., $c_M\approx 0.005$ for $\mu=0.01$) whereas the fraction
of a one-photon state is practically negligible in strong coherent
states. The reduction down to 40~\% can be ideally reached for
such states.
We note that a real source would be characterized
by higher values of the ratio $ w_M $ due to imperfect coupling of
photons in a pair \cite{my}.

It should be mentioned that in fact our measured value of the multi-photon
content is slightly influenced by the noises in the device. The
noise counts may cause multichannel detection even when only
a single photon from the laser pulse has been detected. It is
difficult to determine exactly the contribution of these false
multi-detections. However, it is possible to estimate the upper
limit of false detections due to afterpulses as $p_M^{false} < p_1
p_{ap}^{peak} q $, where $q$ is a duty factor of the detector
channels at the time-of-flight spectrometer (i.e., the ratio of
the channel time window to the time distance between channels;
$q=0.17$ in our case).
Then the contribution to the multi-photon
content due to afterpulses $c_M^{false} < p_{ap}^{peak} q \approx
1.4\times10^{-3}$. For very weak pulse energies false multiple
detections might rather stem from dark counts of the detector.
Nevertheless, the latter would become notable only at
mean-photon-number levels below $10^{-5}$ photon per pulse.
Therefore, the influence of noises is negligible in our device.

The analyzed device might help in the preparation of
photon-number-squeezed states by postselection from entangled
photon pairs obtained by spontaneous parametric downconversion
\cite{my}. Vacuum states can easily be filtered out similarly to
the case of postselection with a single detector \cite{brassard}
and the fraction of multi-photon states can be significantly
reduced using multichannel detection. The obtained states are
potentially useful, e.g., for
quantum-key distribution systems. Our device is a cheap, though
not better as for the performance, alternative to the cascading
detector \cite{kok} for this purpose. Other devices like those
based on NV centers, quantum dots or using detectors with a
superconducting microcalorimeter are quite demanding even for a
common laboratory practice, at least at present.

In this work we use only the information whether single or
multiple detections occurred regardless of the information in
which channels the photons were detected. Since the
channels exhibit different detection probabilities, there is
an additional
information which might be used. In general, it is
possible to obtain the photon-number statistics
of a detected quantum state to some extent. This is a topic of a
forthcoming publication.

In conclusion, we have built a configurable multichannel detector
using a single avalanche photodiode. We have characterized the
best regimes for its operation as multi-photon resolving detector.
Our analysis shows, that this device can efficiently distinguish
multi-photon states from the one-photon state and the vacuum state
over a wide range of energies of the input state.

\section*{Acknowledgments}
The authors acknowledge the support by the projects Research
Center for Optics (LN00A015), CEZJ-14/98 and RN19982003013 supported
by the Ministry of Education of the Czech Republic. They also thank Pavel
Trojek for his contribution to the experimental setup.

\end{document}